\documentclass[prb,preprint]{revtex4}
\usepackage{amsmath}
\usepackage{amsfonts}
\usepackage{graphicx}

\begin{document}

\title{
Polymer Segmental Cross-Correlations from Dielectric Relaxation Spectra of Block Copolymers
}

\author{George D. J. Phillies}
\email{phillies@wpi.edu}

\affiliation{Department of Physics, Worcester Polytechnic
Institute,Worcester, MA 01609}

\begin{abstract}
Dielectric relaxation spectra of block polymers containing sequential type-A dipoles are considered. Spectra of a specific set of block copolymers can be combined to isolate the dynamic cross-correlation between the motions of two distinct parts of the same polymer chain. Unlike past treatments of this problem, no model is assumed for the underlying polymer dynamics.
\end{abstract}

\maketitle

\section{Introduction}

Application of dielectric relaxation to the study of polymer dynamics is readily traced back to the contributions of Stockmayer and Baur\cite{stockmayer1964a,stockmayer1967a}, who treat polar groups in chain molecules, distinguishing between polar group contributions that lie along the chain molecule (type A dipoles), polar group contributions that lie perpendicular to the chain backbone (type B dipoles), and polar group contributions free to rotate internally with respect to the backbone (type C dipoles).  Polymers with type-A dipoles may also have type-B or type-C dipoles.   Type A dipoles are most significant in polymers such as polyesters (-CHR-CO-O-)$_{\rm n}$ that lack a center of inversion.  Up to certain technical issues related to the polymer's detailed chemical structure, the sum ${\bf P}$ of the type-A dipoles is parallel and linearly proportional to the polymer's end-to-end vector ${\bf r}$, allowing dielectric relaxation to be used to study whole-chain polymer dynamics.  The technical issues, in which the backbone contains alternating segments that have and that lack a type A dipole moment, allow ${\bf P}$ and ${\bf r}$ to fluctuate with respect to each other. Extensive studies by Adachi\cite{adachi1993a}, Watanabe\cite{watanabe2001a}, and collaborators demonstrate the wide range of different physical quantities that may be obtained from dielectric measurements, including full dielectric relaxation spectra and principal relaxation times, not to mention a static parameter, the mean-square end-to-end distance $\langle r^{2}\rangle$, and the dependences of all these quantities on polymer molecular weight and concentration.

Adachi, et al.\cite{adachi1991a} have further demonstrated how miscible block copolymers, i.e., block copolymers that do not undergo phase separation, can be used to study dynamics of part of a polymer chain. Two distinct chemical approaches arise.  First, one may form the copolymer  of blocks that are dielectrically active and blocks that are dielectrically inert, the simplest being an AB block copolymer. Measurements on such chains only observe motions of the dielectrically active regions, gaining information on subchain dynamics.  Alas, the active and inert components are chemically different, so their segmental dynamics are in general not the same, complicating interpretation of the results.  Second, as studied by Watanabe, et al.\cite{watanabe1993a},  Urukawa and Watanabe\cite{urukawa1997a}, and others, one may form the copolymer of extended blocks that are chemically identical, but inverted end-to-end, for example (ABC)$_{\rm n}$-(CBA)$_{\rm m}$.  In the second case, all segments have the same dynamic properties. The chemical identity of the two types of block substantially eliminates microphase formation issues that may arise when the blocks are chemically different.  Calculations of the observable dynamics of block copolymers using the Rouse and related models have been made by Tang\cite{tang1996a}, Watanabe, et al.\cite{watanabe1993a} and others.

The objective here is to examine how dielectric studies on either sort of block copolymer can be used to determine cross-correlations between the segmental motions of distinct blocks on a single molecule.  The analysis is fundamentally unlike the other work noted above, in that no assumption is made as to the nature (Rouse, Zimm, reptation, etc.)  of the segmental dynamics.  The proposed approach places substantial demands on the experimental accuracy of the required dielectric measurements, and on the quality of the chemical syntheses of the required polymers, the reward being the determination of elsewise-inaccessible information on polymer dynamics.

For a single chain composed of type-A dipoles, the dipole moment ${\bf P}$ is determined by the dipole moments ${\bf P}_{i}$ of the $N$ individual segmental units via
\begin{equation}
       {\bf P}(t) = \sum_{i=1}^{N} \Theta_{i} {\bf P}_{i}(t).
      \label{eq:Psumform}
\end{equation}
Here the orientation insertion factor $\Theta_{i}$ gives the orientation of the $i^{\rm th}$ segmental unit with respect to the polymer chain, namely $\Theta_{i} = \pm 1$ for a segmental unit inserted into the polymer backbone in the forward or retrograde directions, while $\Theta_{i} =0$ corresponds to the insertion of a neutral segment having zero dipole moment. The assertion that blocks may be inserted in forward or retrograde directions constrains the possible chemical identities of the polymeric units. The dipole-dipole correlation function for a single chain is then
\begin{equation}
       \Phi(t) \equiv \langle {\bf P}(0) \cdot {\bf P}(t) \rangle = \sum_{i, j=1}^{N} \Theta_{i} \Theta_{j} \langle {\bf P}_{i}(0) \cdot {\bf P}_{j}(t) \rangle.
       \label{eq:Phidef}
\end{equation}
In non-dilute solutions, there arises the further possibility that dipoles on adjoining chains have dynamic correlations.  This possibility is realized with solutions of liquid crystalline or rodlike polymers.  Interchain dynamic correlations could readily be incorporated in the discussion via a modest increase in the complexity of the notation, if it were found desirable to do so.  However, for high-molecular-weight flexible chains such correlations are unlikely to be important.

The ${\bf P}_{i}(t)$ for end-to-end type-A dipoles are determined by the positions of the backbone atoms at the ends of each segment.  Locating the ends of dipole $i$ at ${\bf r}_{i}$ and ${\bf r}_{i-1}$, the atoms at the two ends of the polymer are located at ${\bf r}_{0}$ and ${\bf r}_{N}$, while the dipole and segmental vectors are related by
\begin{equation}
      \Theta_{i} {\bf P}_{i}(t) = \Theta_{i} \mu_{0} ({\bf r}_{i}-{\bf r}_{i-1}).
      \label{eq:Praligned}
\end{equation}
If $\theta_{i}=+1$, the head of the dipole is located at ${\bf r}_{i}$, while if $\theta_{i}=-1$, the head of the dipole is located at ${\bf r}_{i-1}$.

For the special case $\Theta_{i} = +1$, $\forall i$,
\begin{equation}
      {\bf P}(t) = \mu_{0} ({\bf r}_{N}(t) -{\bf r}_{0}(t)).
      \label{eq:muparallel}
\end{equation}

The dipole-dipole correlation function may be measured with dielectric relaxation spectroscopy. The review of Williams\cite{williams1972a} shows that $\Phi(t)$ determines the complex dielectric relaxation function $\epsilon^{*}(\imath \omega)$, namely
\begin{equation}
    \epsilon^{*}(\omega) - \epsilon_{\infty} = (\epsilon_{0}-\epsilon_{\infty})\left(1-\imath \omega \int_{0}^{\infty} dt \Phi(t) \exp(-\imath \omega t)\right)
    \label{eq:dielectric1}
\end{equation}
Here $\omega$ is the measurement frequency, and $\epsilon_{0}$ and $\epsilon_{\infty}$ represent the high- and low-frequency limits. The two components $\epsilon'(\omega)$ and $\epsilon''(\omega)$ of the dielectric response are linked by Kronig-Kramers relations, so $\Phi(t)$ may be obtained from either component (if known over a full range of frequencies), via an inverse Fourier transform with its attendant numerical and data-accuracy challenges.

Several cases are of interest.  First, suppose the polymer has only one block, so that $\Theta_{i} = 1$, $\forall i$. Defining the end-to-end vector as ${\bf R}_{2}(t) = {\bf r}_{N}(t) -{\bf r}_{0}(t)$, the dipole correlation function becomes
\begin{equation}
    \Phi_{22}(t) = \mu_{0}^{2} \langle {\bf R}_{2}(0) \cdot {\bf R}_{2}(t) \rangle
    \label{eq:noinversion}
\end{equation}
The dipole-dipole time correlation function is thus the same up to constants as the time correlation function of the polymer end-to-end vector.

Second, suppose that the polymer has a single inversion point, so that $\Theta_{i} = +1$, $i \leq a$, and $\Theta_{i} = -1$, $i \geq a$.  The dipole moment is then proportional to
the vector ${\bf R}_{3}(t) = {\bf r}_{N}(t) +{\bf r}_{0}(t)- 2{\bf r}_{a}(t)$, so the dielectric relaxation function becomes:
\begin{equation}
    \Phi_{33}(t) =  \mu_{0}^{2} \langle {\bf R}_{3}(0) \cdot {\bf R}_{3}(t) \rangle
    \label{eq:inversion}
\end{equation}
A natural choice is $a = N/2$.  The vectors ${\bf R}_{2}(t)$ and ${\bf R}_{3}(t)$ are orthogonal.  Together with the center-of-mass vector ${\bf R}_{\rm cm}$, they are the first elements of an ascending series of orthogonal collective coordinates that can replace the ${\bf r}_{i}$.   Eq.\ \ref{eq:inversion} shows that by examining $\Phi(t)$ for a polymer having a single central point of inversion, the time correlation function of ${\bf R}_{3}$ is directly measurable, and similarly for higher members of the ${\bf R}_{i}$.

Additional information is given by the difference between two $\Phi(t)$.  Suppose we have chains of two species $1$ and $2$ of equal molecular weight, whose orientation insertion factors are denoted $\Theta$ and $\Theta'$, respectively.  Beginning from eq.\ \ref{eq:Phidef}, the difference between the dipole-dipole relaxation functions of the two species is
\begin{equation}
      \Phi_{11}(t) - \Phi_{22}(t) = \sum_{i=1}^{N} \sum_{j=1}^{N} (\Theta_{i}\Theta_{j} -\Theta'_{i}\Theta'_{j}) \langle {\bf P}_{i}(0){\bf P}_{j}(t) \rangle.
      \label{eq:phidifference}
\end{equation}
The correlation functions  $\Phi_{11}(t)$ and $\Phi_{22}(t)$ are measured in separate experiments on different solutions.  There is no implication that cross correlations between the motions of two chains, one of each species, have been measured. The quantity $\Theta_{i}\Theta_{j} -\Theta'_{i}\Theta'_{j}$ functions as a discriminant.  It vanishes if segments $i$ and $j$ have the same orientations in both polymers, or if segments $i$ and $j$ both have opposite orientations in the two polymers.  Taking species 1 and 2 to be the one-block and the single-inversion-point polymers,
\begin{equation}
      \Phi_{11}(t) - \Phi_{22}(t) = 2 \sum_{i=1}^{a} \sum_{j=a+1}^{N} \langle P_{i}(0) P_{j}(t) + P_{i}(t)P_{j}(0)\rangle,
      \label{eq:R2R3}
\end{equation}
which is the time cross-correlation function for the segmental orientations of the two halves of the polymer.  The subtraction process has automatically time-symmetrized the segment-segment time correlation function to include the correlations between segments on either half of the chain at time 0 and on the other half of the chain at time $t$.

Successful spectral subtraction requires careful attention to measuring absolute intensities and normalizations.  The reward for making these demanding measurements is that one may in principal measure the  time-dependent cross-correlation function for the orientations of the two half-chains.  $\Phi_{11}(t) - \Phi_{22}(t)$ is implicitly a distinct correlation function.  It contains no self terms that compare the orientation of the same chain segment at two different times.

Finally, a subtraction process that determines the time-dependent cross-correlation between two shorter segments of two polymer chains is demonstrated. The polymer chains are composed entirely of type-A dipoles.  The two chain segments of interest are labeled block A and block B; they are non-overlapping. Relative to a single-block polymer in which all segmental dipoles point in the same direction, a total of four different polymers are needed, namely the original single-block polymer, two polymers in which block A or block B but not both are inverted, and the polymer in which blocks A and B are both inverted.  Using $A$ and $B$ to denote polymers in which blocks A or B, respectively, are in their initial directions, and using $a$ or $b$ to denote polymers in which blocks A and B, respectively, have been inverted, the dipole relaxation functions of the four polymers are measured, and the difference
\begin{equation}
      \Delta \Phi(t) =  \Phi_{AB}(t) +\Phi_{ab}(t) -\Phi_{Ab}(t) - \Phi_{aB}(t)
      \label{eq:fourwaydelta}
\end{equation}
is determined.
By direct calculation
\begin{equation}
        \Delta \Phi(t) =  4 \sum_{i \in A} \sum_{j \in B} \langle {\bf P}_{i}(0) \cdot {\bf P}_{j}(t)  + {\bf P}_{i}(t) \cdot {\bf P}_{j}(0) \rangle,
        \label{eq:crosscorrelation}
\end{equation}
the notation $i \in A$ and  $j \in B$ denoting sums over all segmental dipoles in Block A and block B, respectively. $\Delta \Phi(t)$ is the time-symmetrized cross-correlation function for the orientations of blocks $A$ and $B$. The time-symmetrization is created by the calculation, and is not imposed {\em post facto} from outside. Implicit in this analysis is the assumption that cross-correlation functions $\langle {\bf P}_{i}(0) \cdot {\bf P}_{j}(t) \rangle$ do not change significantly when blocks $A$ or $B$ are inverted. Such an assumption is appropriate if the two blocks are quite long and well-separated, but will require improvement if the blocks are extremely short and close together.

It has thus been shown that dielectric relaxation spectroscopy can be used to measure the two-time dynamic cross-correlations between the orientations of an arbitrary pair of non-overlapping blocks of a larger polymer. No assumption was made as to a detailed theoretical model, e.g., Kirkwood-Riseman, for describing polymer motions.  Because the calculations are model-independent, analyses based on eqs.\ \ref{eq:R2R3} or \ref{eq:crosscorrelation} can be used to test the validity of particular models for polymer dynamics without being at risk of being criticized for circular reasoning. The analysis here continues to be valid if the two blocks, and their associated type-A dipoles, have been inserted into a polymer whose remaining subunits are dielectrically inert.

\end{document}